\documentclass[11pt,twoside]{article}

\usepackage{asp2006}
\usepackage{epsf}
\usepackage{psfig}
\usepackage{lscape}

\begin{document}
\title{Synchrotron Flaring in Galactic and Extragalactic Jets}   %%% Fill in title
\author{Elina J. Lindfors\altaffilmark{1,}\altaffilmark{2} and Marc T\"urler\altaffilmark{3,}\altaffilmark{4}}   %%% Fill in author names

\altaffiltext{1}{Tuorla Observatory, University of Turku, 21500 Piikki\"o, Finland} \altaffiltext{2}{Mets\"ahovi Radio Observatory, Helsinki University of Technology, 02540 Kylm\"al\"a, Finland}\altaffiltext{3}{INTEGRAL Science Data Centre, ch. d'Ecogia 16, 1290 Versoix, Switzerland} \altaffiltext{4}{Geneva Observatory, University of Geneva, ch. des Maillettes 51, 1290 Sauverny, Switzerland} %%% Fill in author affiliations

\begin{abstract}We study the synchrotron flaring behaviour of the blazar 3C~279 and microquasar Cyg~X-3. The properties of a typical outburst are derived from the observations by decomposing multi-frequency lightcurves into series of
self-similar events. We also discuss the similarities and differences
in flaring behaviour of the galactic and extragalactic jets.
\end{abstract}

\section{Introduction}
High-energy stellar mass binary systems exhibiting jets with apparent
superluminal motions are called microquasars. They are generally
regarded as the galactic counterpart of more powerful extragalactic
sources, quasars. This implies that the underlying physical processes
that operate in these systems are the same and thus are manifested in
their radiative processes.

We have used the shock-in-jet model of Marscher \& Gear (1985), which
describes analytically a shock propagating downstream in the
relativistic jet, to decompose lightcurves of quasars 3C~273 (T\"urler
et al. 1999, 2000) and 3C~279 (Lindfors et al. 2006) and microquasars
GRS~1915+105 (T\"urler et al. 2004) and Cyg~X-3 (Lindfors et al. in
prep.). Here we present some results for 3C~279 and Cyg~X-3. More
information on this method, figures and animation can be found at:
http://isdc.unige.ch/$\sim$turler/jets/.

\section{Results for 3C~279 and Cyg~X-3}
For 3C~279 we find that the lightcurve decomposition, consisting of 10
years of monitoring data from 19 frequencies, supports the Marscher \&
Gear model. The outburst evolution follows the characteristic three
stage pattern but instead of the synchrotron plateau stage, the flux
density is already decreasing during the second stage. %This might be
%due to highly non-adiabatic jet flow, where radiative losses play
%significant role. 
Recently, modifications to Marscher \& Gear model
proposed by Bj\"ornsson \& Aslaken (2000) were implemented to the code
and fitting was redone. Using this model the flux increased during the
second stage, making it difficult to distinguish the synchrotron stage
from the Compton stage. The same was found for 3C~273 (see T\"urler \&
Lindfors 2006).

For the Feb.-Mar. 1994 outbursts of Cyg X-3 (Fig.~1) the average
outburst peaks at $\sim$4 GHz, with a peak flux of 1 Jy, but the scatter of
individual outbursts is large and unlike for GRS~1915+105 they do not
vary only in flux. The decomposition allows us to study the jet
properties and for Cyg~X-3 we find that the electron energy
distribution is rather hard with a slope of 1.68. This is most naturally
explained by turbulence in the shocks. Like for GRS~1915+105, we find
that the jet opening angle increases with the distance.

\begin{figure}
\plotone{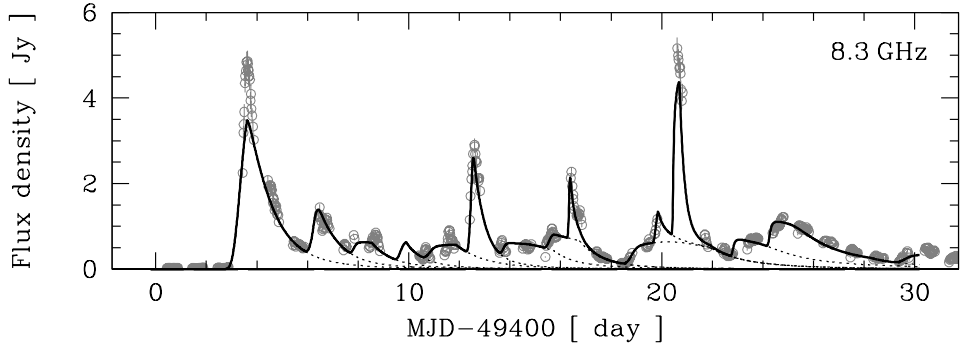}
\caption{One out of four lightcurves of Cyg~X-3 with model outbursts
(dashed lines). Data are from Fender et al. (1997)}
\end{figure}

\section{Discussion}
Similarities that we find among all four sources: 1) Synchrotron
flaring behaviour can be described with a shock-in-jet model. The main
difficulty is to produce the sharp peaks of the outbursts in high
frequencies while the overall shape of the lightcurves is described
well. 2) Non-adiabatic processes seem to play important role and can
not be neglected in the modelling.

Differences between extragalactic and galactic sources in our study:
1) The jets of the galactic sources are apparently widening with
distance (like a trumpet), while quasars tend to have rather conical
jets. This difference might originate from different environment where
the jets form. 2) Recent modelling (T\"urler \& Lindfors 2006)
suggests that the two galactic sources have a harder electron
energy distribution than the two extragalactic sources.

\end{document}